\documentclass{aa}

\usepackage{graphicx}
\usepackage{multirow}

\def\be{\begin{equation}}
\def\ee{\end{equation}}

\catcode`\@=11 
\def\@versim#1#2{\vcenter{\offinterlineskip
        \ialign{$\m@th#1\hfil##\hfil$\crcr#2\crcr\sim\crcr } }}

\begin{document}

\title{Is the energy generation rate of nuclear reactions in hot
  accretion flows important?}

\author{Hui Zhang\inst{1,2}
        \and Yu Wang\inst{3}
        \and Feng Yuan\inst{1}
        \and Fei Ding\inst{3}
        \and Xinlian Luo\inst{4}
        \and Qiuhe Peng\inst{4,5}
}

\institute{ Key Laboratory for Research in Galaxies and Cosmology,
Shanghai Astronomical Observatories, Chinese Academy of Sciences, 80
Nandan Road, Shanghai 200030, China; hzhang@shao.ac.cn,
fyuan@shao.ac.cn
    \and
Graduate School of the Chinese Academy of Sciences, Beijing 100039,
China
    \and
University of Science and Technology of China, Hefei, Anhui, 230026,
China; ywa@mail.ustc.edu.cn
    \and
Department of Astronomy, Nanjing University, Nanjing, 210093, China
    \and
The Open Laboratory of Cosmic Ray and High Energy Astrophysics,
Chinese Academy of Sciences, Beijing, 100049, China }

\date{Received 00.00.0000 / Accepted 00.00.0000}

\abstract{ The temperature of hot accretion flows around black holes
is sufficiently high for the ignition of nuclear reactions. This is
potentially an important nucleosynthesis mechanism in the universe.
As the first step in studying this problem, we need to measure
physical quantities such as density and temperature of the accretion
flow. In usual studies of the hot accretion flow, viscous
dissipation is considered to be the only heating mechanism, while
the heating caused by nuclear reactions is not considered. In this
paper, we investigate whether the energy generation rate of nuclear
reaction is important compared to the viscous heating. Our
calculation indicates that the former is at most one percent of the
latter and thus is not important. The dynamics of accretion flow can
be therefore calculated in the usual way, without the need to
consider heating due to nuclear reactions.

\keywords{Nuclear Reaction, Energy Generation Rate, ADAF} }

\maketitle

\section{ Introduction}
Observations imply that there maybe a supermassive black hole at the
center of almost every galaxies. There is a large fraction for which
the accretion flow can be described by a hot accretion flow, such as
advection-dominated accretion flow (ADAF; Narayan \& Yi 1994, 1995;
Abramowicz et al. 1995; for a review of ADAF, see Narayan, Mahadevan
\& Quataert 1998; for a review of the applications of ADAF, see Yuan
2007) and luminous hot accretion flow (LHAF; Yuan 2001). Hot
accretion flow also exists in some states of black hole X-ray
binaries. The temperature of the hot flow is very high, $
T_{i}\sim10^{12}$K$/r$, $r$ being the radius in unit of
Schwarzschild radius $r_s$. This temperature is high enough to
ignite significant nuclear reactions in the hot flow. This process
is potentially interesting because it is a promising means of
producing new elements. The analysis of observations by Juarez et
al. (2009; see also Hamann \& Ferland 1999) found that the
metallicity of the BLR of redshift quasars is very high. One
explanation is that the massive stars had formed before the quasars
were created, and these stars then enriched the metallicities of
quasars. It is also possible that BLR gas originates in accretion
flows. Although accretion flows in quasars are generally assumed to
be in the regime of higher accretion rates than that of a hot
accretion flow, i.e., the standard thin disk, the detailed structure
of the accretion flow is still largely unknown because the thin disk
can not explain the origin of the X-ray emission widely observed in
quasars. The presence of X-ray emission implies that ADAF or
LHAF-like hot accretion flow is also likely to exist in quasars, not
only in some low-luminosity AGNs. One possible reason for the high
metallicity in quasars is therefore that nucleosynthesis occurs in
the accretion flow (see the review paper of Hamann \& Ferland 1999).

Some people have already studied the nuclear reactions and
nucleosynthesis in accretion flows around black holes (e.g., Jin et
al. 1989, Arai \& Hashimoto 1992, Chakrabarti \& Mukhopadhyay 1999,
Mukhopadhyay \& Chakrabarti 2000, Mukhopadhyay \& Chakrabarti 2001,
Hu \& Peng 2008). In these works, however, the accretion flow models
adopted are quite different from ADAFs. The works of Jin et al. and
Arai \& Hashimoto are based on a ``thick disk'', while the two works
by Mukhopadhyay \& Chakrabarti are based on a simplified inviscid
accretion flow, which differs again from ADAFs. These particular
accretion models have received little attention these years. Hu \&
Peng (2008) studied nucleosynthesis in ADAFs, but their work was
based on the self-similar solution of ADAF. This solution is
affected by significant error in the inner region of the accretion
flow, where most nuclear reactions occur.

We therefore revisit the study of nucleosynthesis in hot accretion
flows. We first constrain the dynamics of accretion flows. Although
the global solution of ADAF has been known for many years, since
nuclear reactions have not previously been taken into account, we
need to investigate whether they are important compared to the
viscous heating. If they are then, the self-consistent global
solution of ADAF needs to be recalculated. We neglect the
photodisintegration because the optical depth in ADAFs is very small
and most of the photos can escape from the hot accretion flow
without being scattered.

The structure of the paper is as follows. We briefly introduce ADAFs
in Sect. 2.1 and the calculation method of energy generation rate in
Sect. 2.2. The calculation results will be shown in Sect. 2.3. In
Sect. 3 we summarize our results.

\section{Energy generation rate in hot accretion flow}

\subsection{Advection dominated accretion flow}

The dynamics of the ADAF is described by the continuity equation,
the radial and azimuthal components of momentum equation, and the
energy equations of electrons and ions (Narayan, Mahadevan \&
Quataert 1998)

\be
\dot{M}=-4\pi R H \rho v = \dot{M}_{\rm out}\left(\frac{R}{R_{\rm out}} \right)^s,
\ee

\be
v \frac{dv}{dr}= -\Omega_{\rm k}^2 r+\Omega^2
r-\frac{1}{\rho}\frac {dp}{dr},
\ee

\be v(\Omega r^2-j)= \alpha r
\frac{p}{\rho},
\ee

\be
\rho v \left(\frac{d\varepsilon_e}{dr}- {p_e \over \rho^2} \frac{d \rho}{dr}\right) =\delta
q^++q_{ie}-q^-,
\ee

\be \rho v \left(\frac{d\varepsilon_i}{dr}- {p_i \over \rho^2} \frac{d
  \rho}{dr} \right) =(1-\delta)q^+-q_{ie},  \ee
where $\dot{M}_{\rm  out}$ is the mass-flow rate at the outer
boundary ($R_{\rm out}$) of the flow, the exponent $s$ describes the
strength of the outflow, $\delta$ is the fraction of turbulent
viscous energy that directly heats the electrons, $q^+$ is the
viscosity heating rate, $q_{ie}$ is the Coulomb energy exchange rate
between electrons and ions, and $q^-$ is the radiative cooling rate
per unit volume. The main radiative processes are synchrotron
emission, bremsstrahlung, and their respective Comptonizations. The
eigenvalue $j$ corresponds to the specific angular momentum loss at
the inner boundary, $\varepsilon_i$ ($\varepsilon_e$) is the
internal energy of ions (electrons), and $P_i$ ($P_e$) is the
pressure due to ions (electrons).

The global solution of these equations needs to satisfy three
conditions. The first one is the no-torque condition at the horizon.
This condition can be automatically satisfied if we adopt the
viscous description of $\tau_{r \phi}=\alpha (P_{gas}+P_{mag})$,
where $\tau_{r \phi}$ is the viscous stress tensor, and $P_{gas}$
and $P_{mag}$ are the gas pressure and magnetic pressure,
respectively. The second condition is the transonic condition. When
the matter enters into the horizon of the black hole, its speed
should be equal to the light speed. A sonic point must therefore
exist where the radial speed of flow equals the local sonic speed.
The solution should finally satisfy the boundary condition at
$R_{out}$. Here we chose the temperature of ions and electrons
$T_{i,e}$, and the ratio of the radial velocity of the flows to the
local speed of sound $v/c_s$ at $R_{out}$. We adopt the Paczy\'nski
\& Wiita (1980) potential to mimic the geometry of a Schwarzschild
black hole. We adjust the eigenvalue $j$ and use the shooting method
to derive the global solution that satisfies the conditions above.
For the parameters of the hot accretion flow, we choose their
typical values constrained by modeling black hole X-ray binaries
(e.g., Yuan, Cui \& Narayan 2005) and the supermassive black hole in
our Galactic center, Sgr A*, which is the most well studied
supermassive black hole system observationally (Yuan, Quataert \&
Narayan 2003). They are the viscous parameter $\alpha=0.3$ (but see
discussions below), $\beta=0.9$, $s=0.3$, and $\delta=0.5$, where
$\beta$ is the ratio of the gas pressure to the total pressure, $s$
describes the strength of the outflow, and $\delta$ is the fraction
of viscous dissipation that directly heats the electrons. The
detailed numerical approach can be found in Yuan (2001) or Yuan et
al. (2003).

 As is well known, when the accretion rate is very low,
almost all of the viscously dissipated energy is stored in the
accretion flow and advected into the horizon of the black hole,
rather than radiated away. This is because the radiative timescale
is far longer than the accretion timescale. However, when the
accretion rate increases, the radiative timescale becomes shorter,
and less energy is advected into the black hole. In the regime of
LHAF, the energy advection even becomes negative. In this case, the
radiative efficiency of the hot accretion flow increases, even
reaching that of the standard thin disk (Shakura \& Sunyaev 1973).
Unlike the standard thin disk, ADAF is of course, still hot and
geometrically thick. The temperature of the ions is almost always
virial, and is a power-law function of radius.

\subsection{Calculation of nuclear reaction heating}

We make the following assumptions. The first is that the element
abundance in the accretion flow is the same as that in the solar
atmosphere (Grevesse \& Sauval 2000). The companion of the stellar
mass black hole could be a main-sequence star, and the accretion
matter of a supermassive black hole at the galactic center is
presumably supplied by a number of nearby stars. Second, we assume
that the element abundance due to nuclear reaction does not differ
significantly from the initial abundance. This is obviously not a
robust assumption if significant nucleosynthesis does occur in the
accretion flow. Before we know the final result, we must however
make this assumption. We could complete some ``iteration'' if we
knew how the abundance changes. Third, in the large-scale universe,
the relative abundance of an isotope compared to other isotopes of
the same element is quite constant. You can find the relative
abundance of isotopes from the table at the website
\footnote{http://www.chem.queensu.ca/FACILITIES/NMR/nmr/mass-spec/mstable3.htm}.
In our calculation here, we accept the assume isotope abundances
given in this table.

The nuclear reactions in the star and in the hot accretion flow are
different. In the star, different sets of burnings are well
separated by appreciable temperature differences in the star.
However, in the hot accretion flow, these nuclear reactions can be
ignited almost simultaneously because the temperature is
sufficiently high to ignite different nuclear burning at the same
time. In principle, nuclear reaction can proceed for many kinds of
elements if the temperature is high enough. We considered only H,
He, and C burning after finding that other nuclear reactions were
not important contributors to the energy generation rate. Our second
assumption is therefore justified.

The temperature of the accretion flow increases when it is closer to
the central black hole, when an increasing number of elements are
burned. Different nuclear reactions are ignited at different
temperatures. Hydrogen is burned when the temperature is higher than
$7 \times 10^6$ K. Helium can be burned when the temperature reaches
$1-2 \times 10^8$ K, and for carbon, this temperature is about $7
\times 10^8$ K. The total energy generation rates in different
temperature ranges are as follows:
\begin{equation}
\epsilon_{nuc}=\left\{ \begin{array}{ll} \epsilon_{H}, & \quad T\geq 7
  \times 10^6K , \\ \epsilon_{H}+\epsilon_{He} , & \quad T\geq 2
  \times 10^8K , \\ \epsilon_{H}+\epsilon_{He}+\epsilon_{C}, & \quad T
  \geq 7 \times 10^8K,
\end{array} \right.   \\
\end{equation}
$\epsilon_{H}$, $\epsilon_{He}$, and $\epsilon_{C}$ is the total
energy generation rate of hydrogen, helium, and carbon,
respectively, and $\epsilon_{nuc}$ represents their sum. They are
defined to be the energy produced per unit mass and time
\begin{equation}
\epsilon=\frac{1}{\rho}\sum r_{ij}Q_{ij},
\end{equation}
where $r_{ij}$ is the reaction rate per unit i.e., volume and time
between the nuclei $i$ and $j$, and $Q_{ij}$ is the energy released
per reaction,

\begin{equation}
r_{ij}=\frac{1}{1+\delta_{ij}}n_in_j \langle \sigma v \rangle _{ij},
\end{equation}
where,
\begin{equation}
\delta_{ij}=\{ \begin{array}{ll} 0, &i \neq j, \\
1, &i = j,\end{array}
\end{equation}
$\langle \sigma v \rangle _{ij}$ is the average cross-section, and
$n_i$ and $n_j$ is the number density of particles $i$ and $j$,
which can also be written as
\begin{equation}
n_{i}=\rho N_{A} \frac{X_i}{A_i}=\rho N_{A}Y_i,
\end{equation}
where $X_i$, $Y_i$, and $A_i$ are the mass abundance, number
abundance, and atomic weight of element $i$, respectively, and $N_A$
is Avogadro number. According to Eqs. (7), (8), (9) and (10), the
energy generation rate of nuclear reactions can be written as
\begin{equation}
\epsilon=\rho N_A \sum \frac{1}{1+\delta_{ij}} Y_iY_jQ_{ij}N_A
\langle \sigma v \rangle _{ij}.
\end{equation}
The detailed calculations can be found in Huang (1998), Fowler et
al. (1975), Harris et al. (1983), and Georgeanne \& Fowler (1988).

We study the energy generation rates of nuclear reactions for
accretion flows of various parameters. For the mass of the black
hole, we consider a stellar mass black hole ($10M_{\odot}$), an
intermediate-mass black hole ($10^3M_{\odot}$), and a supermassive
black hole ($10^8M_{\odot}$). These values are typical of black hole
X-ray binaries, some possible ultraluminous X-ray sources, and AGNs,
respectively. For the case of supermassive black holes, we consider
various mass accretion rates, namely $10^{-4}\dot M_E$ ($\dot M_E$
is the Eddington accretion rate, $\dot M_E=10 L_{Edd}/c^2$),
$10^{-3}\dot M_E$, 0.01$\dot M_E$, and 0.1$\dot M_E$.

After deriving the global solutions of the hot accretion flow by
solving Eqs. (1-5), we measure the physical quantities of the hot
accretion flow as a function of radius in each case, including
temperature $T_i$ and $T_e$ and mass density $\rho$, and calculate
the viscosity heating rate $q^+$ (defined as the viscosity heating
rate per unit volume). We then calculate the energy generation rate
$\epsilon_{nuc}$ of the nuclear reactions at different radii based
on temperatures and densities obtained above from Eq. (11). We note
that $\epsilon_{nuc}$ is the energy generation rate per unit mass,
but $q^+$ is usually defined as the viscosity heating rate per unit
volume. We therefore define $\varepsilon_{nuc}=\rho \epsilon_{nuc}$
for consistency, and also denote $q^+$ by $\varepsilon_{vis}$.
Finally, we derive the ratio of energy generation rate of nuclear
reaction to viscosity heating rate, i.e.,
$\varepsilon_{nuc}/\varepsilon_{vis}$. If this ratio is high, the
ADAF model must obviously be revised by including the heating caused
by nuclear reactions.

\begin{figure}[ht]
\includegraphics[width=75mm,height=75mm]{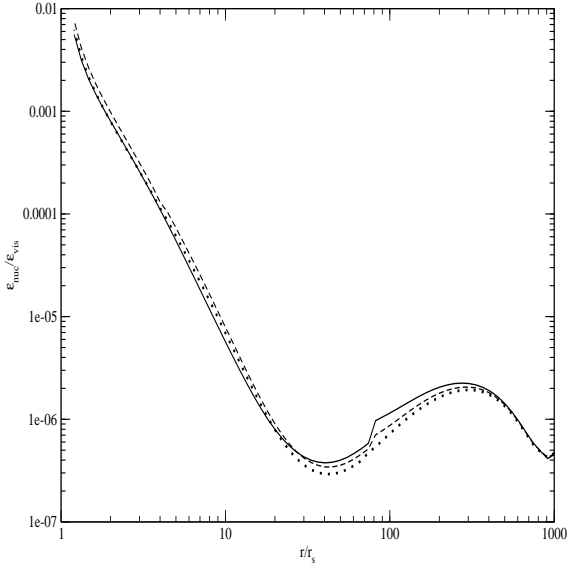}
\label{fig:2} \caption{The ratio of nuclear-reaction energy
generation rate to viscous heating
  rate as a function of radius for different black hole masses. The solid,
  dashed, and dotted lines represent the cases of $M=10M_{\odot}$,
  $10^3M_{\odot}$, and $10^8M_{\odot}$. The mass accretion rate
  $\dot{M}= 0.1\dot M_E$ in all three cases.}
\end{figure}

\begin{figure}[ht]
\includegraphics[width=75mm,height=75mm]{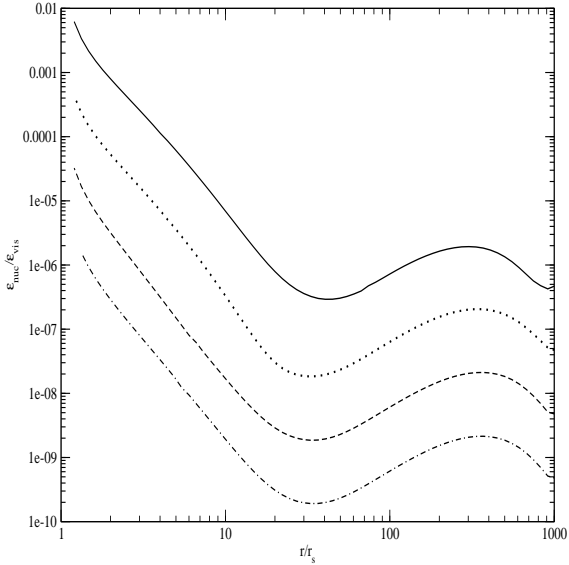}
\label{fig:3} \caption{The ratio of
  energy generation rate of nuclear reaction to the viscous heating
  rate as a function of radius for different mass accretion rates. The
  solid, dotted, dashed, and dot-dashed lines represent the cases of
  $\dot{M}=0.1\dot M_E$, $10^{-2}\dot M_E$, $10^{-3}\dot M_E$, and $10^{-4}\dot M_E$.
  The mass of the black hole is $10^8M_{\odot}$.}
\end{figure}

\subsection{Calculation results}

The calculation results are shown in Figs. 1 \& 2. Figure 1 shows
the ratio of the heating rate for nuclear reactions to that for
viscous dissipation for $\dot{M}=0.1\dot{M}_{\rm E}$ but different
black hole masses. We find from this figure that the ratio is
roughly proportional to $(r/r_s)^{-3}$. This is because in our case
$\varepsilon_{nuc}$ is roughly proportional to $\rho^2 T^5$, while
$\varepsilon_{vis}\propto \rho r^{-5/2-s}$. So
$\varepsilon_{nuc}/\varepsilon_{vis} \propto \rho r^{5/2+s} T^5$.
Since $\rho\propto (r/r_s)^{-3/2+s}$ while $T\propto r^{-1}$, we
have $\varepsilon_{nuc}/\varepsilon_{vis}\propto r^{2s-4}\propto
r^{-3.4}$. The ratio does not depend on the mass of the black hole.

Figure 2 shows this ratio as a function of radius for different
$\dot{M}$ but the same black hole mass. The ratio is roughly
proportional to $\dot{M}$ (or density), because the nuclear reaction
heating rate is proportional to $\dot{M}^2$ while the viscous
heating is proportional to $\dot{M}$, and the ions temperature is
always virial, independent of $\dot{M}$. However, even at
$\dot{M}=0.1\dot{M}_{\rm E}$, almost the highest accretion rate a
hot accretion flow can have, this ratio is less than $1\%$, i.e.,
the heating rate of nuclear reactions in the hot accretion flow is
not important. We therefore do not need to include the nuclear
reaction in calculating the dynamics of ADAF.

In our calculation, the value of the viscosity parameter $\alpha$ is
adopted to be 0.3, which is ``typical'' of ADAFs. We should
emphasize that this value has some uncertainties. The value of
$\alpha$ is perhaps constrained most reliably by observations of
dwarf nova outbursts, where $\alpha \sim 0.1-0.2$ (Smak 1984). Three
dimensional magnetohydrodynamic (MHD) numerical simulation suggest
that $\alpha \sim 10^{-2}-10^{-1}$ (Hawley \& Krolik 2001). King et
al. (2007) argued that according to the highest quality
observational evidence, $\alpha \sim 0.1 -0.4$ in thin and fully
ionized discs. For a given accretion rate, the density of the flow
is proportional to $\alpha^{-1}$. The ratio of the heating rate
between nuclear reaction and viscous dissipation is therefore
proportional to $\alpha^{-1}$.

\section{Summary and discussion}

The temperature of the hot accretion flow such as ADAFs is very
high, so nuclear reaction should occur. To study the
nucleosynthesis, the first step is to determine the dynamics of the
accretion flow. For this purpose, we have investigated the ratio of
the heating rate for the nuclear reactions to that for the viscous
dissipation. We have found that the ratio increases with accretion
rates. However, even for the highest accretion rate for which the
hot accretion flow solution exists, the ratio is $\la$ 1\%.
Therefore, the heating caused by nuclear reaction can be neglected.
This result indicates that the previously obtained global solution
of hot accretion flow is self-consistent and can be used directly in
future studies of nucleosynthesis.

As we mentioned in introduction, some authors have previously
studied nuclear reactions and nucleosynthesis in accretion flows
around black holes (e.g., Jin et al. 1989, Arai \& Hashimoto 1992,
Chakrabarti \& Mukhopadhyay 1999, Mukhopadhyay \& Chakrabarti 2000,
Mukhopadhyay \& Chakrabarti 2001). They found that significant
nuclear reactions can occur and many heavy elements can be
synthesized, which is quite different from our results. There are
three reasons for this discrepancy. The first is that, as we
mentioned in the introduction, the works by Jin et al. and Arai \&
Hashimoto are based on a ``thick disk'' model in which the radial
velocity is several orders of magnitude lower than that in ADAFs if
the accretion rate is the same. Secondly, the accretion rates in
most of the above-mentioned works are super-Eddington, while the
accretion rate in ADAFs is lower than the Eddington rate. The third
reason is that the value of the viscous parameter adopted in
previous work is much smaller than that we adopt, $\alpha=10^{-7}$
(Jin et al. 1989), $10^{-10}$ (Arai \& Hashimoto 1992), and $0$
(Chakrabarti \& Mukhopadhyay 1999; Mukhopadhyay \& Chakrabarti 2000;
Mukhopadhyay \& Chakrabarti 2001), respectively. So the density of
the accretion flow will be much higher than that in ADAFs. Since the
efficiency of nuclear synthesis will increase in proportion to
density, this explains why they found that the nuclear reaction is
so significant.


\begin{acknowledgements}
We thank Lei Zhao for useful discussions on nuclear reaction. This
work was supported in part by the Natural Science Foundation of
China (grants 10773024, 10833002, 10821302, and 10825314), Bairen
Program of Chinese Academy of Sciences, and the National Basic
Research Program of China (973 Program 2009CB824800).
\end{acknowledgements}

\clearpage
\end{document}